# Nanoscale Intergranular Corrosion and Relation With Grain Boundary Character as Studied *In Situ* on Copper

*Mohamed Bettayeb,[a] Vincent Maurice,[a,\*] Lorena H. Klein,[a] Linsey Lapeire,[b] Kim Verbeken,[b] and Philippe Marcus[a,\*]*

[a] PSL Research University, CNRS - Chimie ParisTech, Institut de Recherche de Chimie Paris (IRCP), Physical Chemistry of Surfaces Group, 11 rue Pierre et Marie Curie, 75005 Paris, France

[b] Department of Materials, Textiles and Chemical Engineering, Ghent University (UGent), Technologiepark 903, 9052 Zwijnaarde (Ghent), Belgium

## Abstract

The initiation of intergranular corrosion at various types of grain boundaries (GBs) was studied at the nanometer scale on microcrystalline copper in 1 mM HCl aqueous solution. *In situ* Electrochemical Scanning Tunneling Microscopy (ECSTM) and Electron Back-Scatter Diffraction analysis of the same local microstructural region were combined using an innovative methodology including micro marking performed with the STM tip. The results demonstrate that electrochemically-induced intergranular dissolution, at the surface termination of GBs, is dependent on the grain boundary character. It is found that random high angle boundaries as well as $\Sigma 9$ coincidence site lattice (CSL) boundaries are susceptible to nanoscale initiation of intergranular corrosion while for $\Sigma 3$ CSL boundaries the behavior is dependent on the deviation angle of the GB plane from the exact orientation. For the $\Sigma 3$ twins, a transition from resistance to susceptibility occurs between 1° and 1.7° of deviation as a result of the increase of the density of steps (i.e. misorientation dislocations) in the coincidence boundary plane. The work emphasizes the precision needed in the design of the grain boundary network in applications where intergranular corrosion or its initiation must be controlled at the nanoscale.

Keywords: Copper; STM; Nanometer scale; Intergranular corrosion; Grain boundary structure

[\*] Corresponding authors:
V. Maurice (vincent.maurice@chimie-paristech.fr); P. Marcus (philippe.marcus@chimie-paristech.fr)

# 1. Introduction

Most technical metallic materials, including the newly developed microcrystalline and nanocrystalline metals with specific properties [1-4], are polycrystals and expose at their surface an extensive amount of grains joined by a grain boundary (GB) network. Their durability in aggressive environments is sometimes limited by intergranular corrosion that degrades the GB network first at the surface and then penetrates in the sub-surface and propagates to jeopardize the integrity of the entire microstructure. Grain boundary engineering aims at improving the behavior of polycrystalline materials towards phenomena such as intergranular corrosion, sensitization and crack propagation, by controlling the GB network characteristics. This is a challenging task in terms of process development but also scientifically since the design of the most resistive GB network requires an improved understanding of the relation between GB character and the local properties of corrosion resistance at the surface sites where the grain boundaries emerge, including in the initial stages of corrosion before penetration in the sub-surface region and propagation.

Previous studies have discussed how intergranular corrosion relates to the GB character and GB energy, based on micrographic and microscopic characterization of GB engineered materials submitted to intergranular sub-surface penetrating attack [5-21]. High angle grain boundaries are classified as more reactive because of their higher energy compared to low angle GBs. Low angle GBs can be described by a network of edge dislocations (or steps) in the GB plane for a misorientation angle not exceeding 15°. Among the high angle GBs, some can be described by a coincidence site lattice (CSL) and labeled $\Sigma n$ with $1/n$ defining the fraction of lattice sites in common in the two grains. These CSL grain boundaries, especially the low $\Sigma$ ones, i.e. those with a high fraction of common lattice sites, have lower energies and are so-called "special" when they resist degradation as opposed to the high Σ ones and random high angle grain boundaries [6,8,10,13-15,17]. Among the low Σ boundaries, there are the Σ3 boundaries or twins which are most common in face centered cubic materials such as copper. It has been reported that the Σ3 and only partially the Σ9 boundaries would resist to intergranular corrosion [8,11,14,15,17]. Moreover, the Σ3 twins can be coherent or incoherent depending on the specific orientation of the GB plane and it has been reported that only the Σ3 coherent twins (with a {111}-oriented GB plane) would better resist intergranular corrosion [7,15].

Electro-Chemical Scanning Tunneling Microscopy (ECSTM) allows studying, locally and with high resolution, *in situ* electrochemically-induced alterations of the topmost surface of metallic materials. On copper, it has been classically applied to single-crystal surfaces by several groups in order to study corrosion and passivation in various electrolytes at the nanometer and atomic scales [22-36]. More recently, ECSTM was applied to microcrystalline copper to investigate *in situ* local active dissolution and local passivation properties at the surface termination of grain boundaries in metals [37-40]. Dissolution in the active state [37] as well as transient dissolution during passivation [38,40] were found weaker at the GB edges assigned to coherent twin boundaries than at other GB edges assigned to random or other CSL boundaries. These studies were, however, limited by the absence of detailed characterization of the GB character of the local sites where corrosion could be measured at the nanoscale by ECSTM.

Here, we report developments by which we are able to combine local topographic analysis at the nanoscale at the surface termination of grain boundaries by ECSTM under electrochemical control with local microstructural analysis of the GB character by Electron Back-Scatter diffraction (EBSD). An innovative micro marking method has been developed to allow repositioning and measurement of the same local area with both techniques. This combined analytical approach was applied, for the first time to our knowledge, for a comparative study of the initial stages of intergranular corrosion in the active state at and along various types of grain boundaries.

## 2. Experimental

Samples of microcrystalline copper obtained by cryogenic rolling of high-purity cast electrolytic tough pitch (ETP-) Cu were used [20,37-42]. Annealing was limited to 1 min at 200 °C to ensure full recrystallization while preserving a suitable grain size and grain boundary density for the STM field of view as confirmed by EBSD [37,38]. No preferential grain orientation, i.e. a nearly random texture, was obtained [38] and 66% of the GB length corresponded to $\sum 3$ CSL boundaries, the other 34% being random boundaries for the most part [37,38]. Mechanical polishing with diamond spray down to a final grade of 0.25 µm was applied for surface preparation, followed by electrochemical polishing in 66% orthophosphoric acid at 3 V for 15 s versus a copper counter electrode in order to remove the cold work layer.

ECSTM analysis was performed with an Agilent Technologies system (PicoSPM base, STM S scanner, PicoScan 2100 controller, PicoStat bi-potentiostat and Picoscan software). The ECSTM cell, its cleaning and tip preparation have been detailed elsewhere [43-47]. Two Pt wires served as pseudo reference electrode and counter electrode and a Viton O-ring delimited a working electrode area of 0.16 cm$^2$. All potentials reported hereafter are relative to the standard hydrogen electrode (V/SHE). Tips were prepared from 0.25 mm diameter tungsten wire by electrochemical etching and covered with Apiezon wax.

The electrolyte was a non-deaerated 1 mM HCl(aq) aqueous solution (pH 3) prepared from ultrapure HCl and Millipore water (resistivity > 18 MΩ cm). In order to avoid any uncontrolled dissolution, the samples were exposed to the electrolyte at -0.35 V/SHE (open-circuit potential value of - 0.25 V/SHE). Then the potential was repeatedly cycled (0.2 V s$^{-1}$) down to - 0.68 V/SHE at the onset of hydrogen evolution and backward to the value of - 0.45 V/SHE until the cyclic voltammograms (CVs) indicated that no cathodic peak associated with Cu(I) to Cu(0) reduction was observed. Two cycles were enough to fully reduce the air-formed native oxide.

After this cathodic reduction pre-treatment, images of the microcrystalline copper surface in the metallic state were taken at -0.55 V/SHE for localization of grain boundaries of interest. Afterwards, anodic dissolution was forced by cycling (0.2 V s$^{-1}$) the potential first up to an anodic apex at a current density of about 30 to 50 µA cm$^{-2}$ then backward to the cathodic apex of -0.68 V/SHE and finally upward to the initial value of -0.55 V/SHE. This step was repeated several times and new ECSTM images of the surface were taken at this potential every two cycles after up to 6 dissolution/redeposition cycles. During electrochemical treatment, the STM tip was kept engaged but not scanning the surface, so as to record the very same area of interest after cycling. The local evolution of the topography could thus be followed after electrochemical cycling. The STM images were acquired in the constant current mode. No filtering was used and the recorded images were processed with the Gwyddion software [48].

At the end of the ECSTM experiments, the STM tungsten tip was used to make micro marks on the surface so as to enable precise repositioning for local analysis of the same area of interest by EBSD. After recording the last STM image, scanning was interrupted from the surface but the tip was kept engaged. Using the X and Y piezo scanners, the tip was positioned at one of the four corners of the STM field of view ($10 \times 10$ µm$^2$). Then, the stepper motor of the microscope controlling the Z displacement of the sample was activated so as to indent the surface with the tip. This was done by displacing the sample 1 to 2 µm towards the tip. Next, the tip was displaced using the full range (10 µm) of the X and Y piezo scanners while kept indented in the surface, thus producing a micro mark on the surface in the vicinity of the local area previously analyzed by ECSTM. The micro marks thus made had the form of two perpendicular segments about 10 µm long each and joining at one end. Before transfer of the sample to EBSD analysis, their formation was checked by optical micrographic observation.

EBSD analysis was performed at LISE (Laboratoire Interfaces et Systèmes Electrochimiques, CNRS –Université Pierre et Marie Curie) with a digital scanning electron microscope Ultra55 from ZEISS (FEG-SEM). Secondary electron images were first recorded at an accelerating voltage of 20 kV, a nominal beam size of 10 nm and a measured beam current of 1.5 nA in order to locate the micro marks produced with the STM tip. Then, the analysis in EBSD mode was performed on areas of $33 \times 33$ µm$^2$ and including the micro marks. It was performed with a minimum step size of 0.1 µm in order to optimize the post-processing for the crystallographic characterization of the grain boundaries, including at the local sites where line profile analysis could be performed from the STM data. The EBSD data were processed with the OIM® software.

## 3. Results and discussion

*3.1. Macroscopic electrochemical behavior and relation with GB network*

The macroscopic electrochemical behavior of microcrystalline copper in 1 mM HCl(aq) was characterized by cyclic voltammetry in the ECSTM cell (Figure 1). Figure 1(a) shows a CV started at -0.35 V/SHE in the metallic state obtained after reduction of the air-formed native oxide. For E > -0.25 V/SHE, the increase of the anodic current is assigned to the anodic copper dissolution reaction ($Cu(0) \rightarrow Cu(I) + e^-$) since no stable copper oxide is formed in these acidic conditions (pH 3) [49,50]. In the reverse scan, a cathodic peak is observed at -0.17 V/SHE. It is assigned to the reductive deposition of dissolved copper as observed on single crystals [22,23,26,28,30,36]. At lower potentials, a cathodic peak is observed at -0.56 V/SHE in the potential range for adsorption/desorption of chlorides, as previously discussed for single crystals [22,23,26,28,30,36].

Figure 1(b) shows a typical CV of dissolution/redeposition (-0.40 < E < -0.15 V/SHE). The analysis of the charge density transfer was performed for the anodic and reverse cathodic reactions by integrating the current density measured above and below the drawn baseline, respectively. After 2 CVs, the measured cumulated charge densities were 110±7 $\mu C$ cm$^{-2}$ and 101±7 $\mu C$ cm$^{-2}$ for the anodic and cathodic reactions, respectively, suggesting a quasi-reversible behavior at this stage within the precision of the measurement. After 4 CVs, they were 333±15 $\mu C$ cm$^{-2}$ and 281±15 $\mu C$ cm$^{-2}$, respectively, showing that the process became partially irreversible. Using Eq. (1), where $V_m$ is the molar volume of metallic copper (7.1 cm$^3$ mol$^{-1}$), $z$ the number of exchanged electrons (1), and $F$ the Faraday's constant, the equivalent thickness $\delta$ of reacting copper can be calculated from the measured electric charge densities $q$.

$$\delta = \frac{qV_m}{zF} \qquad \text{Eq. (1)}$$

One obtains values of 0.08 and 0.24 nm after 2 and 4 cycles, respectively. After 4 cycles, this amounts to about 1 equivalent monolayer (ML) of copper being reversibly dissolved (one (111)-oriented ML of copper is 0.208 nm thick from the bulk *fcc* structure).

These CV measurements characterize the macroscopic behavior of the surface. Generally, they can be associated with the properties of the grains since the exposed surface fraction essentially corresponds to grains. However, the reacting quantities are extremely small since in average about 0.06 nm of copper, i.e. ~0.3 ML, reacts for each cycle. Based on this sub-monolayer amount of reacting material, let us assume that the intergranular response would dominate the surface reactivity in these initial stages of corrosion. The reactivity of specific surface regions consisting of the GB edges, i.e. the termination of the GB planes, and their immediate vicinity (hereafter referred as the GB regions) would not be negligible and may even dominate the electrochemical response. Taking into account the average grain size of 1.4 μm of the microcrystalline copper samples, as estimated from the EBSD data [37,38], and assuming an average width of 40 nm for the GB regions at the surface as observed by ECSTM [37,39,40] (i.e. at least one order of magnitude larger than the GB width in the bulk), one can estimate the surface fraction of the GB regions. Assuming for simplicity a square GB network, one finds the surface fraction of the GB regions to be 5.6%. Consequently, an average thickness of about 1.1 nm of copper would dissolve and redeposit in the GB regions at each cycle if one assumes that the grains remain inert and 100% of the GB network react. Further taking into account the length fraction of the high angle random grain boundaries (about 34% for these samples), the thickness of copper reversibly dissolved in the GB regions at each cycle would be about 3.1 nm if only the random grain boundaries would react and the CSL boundaries would remain inert.

These calculations suggest that the applied CV treatments are appropriate to investigate the initial stages of intergranular corrosion, as confirmed by the data discussed below. They also emphasize the nanoscale character of the studied process since the estimated reacting quantities are of the order of nanometers in thickness and must be studied locally at the topmost surface, i.e. before penetration of the corrosive attack in the GB network.

*3.2. Corrosion behavior of Σ3 grain boundaries*

Figure 2 shows an example of a local area of microcrystalline copper where the microstructure was first imaged *in situ* by ECSTM at the topmost surface (Figure 2 (a)) and then mapped by EBSD (Figure 2(b,c)) after sample transfer and repositioning of the analysis using the procedure described above. The micro mark done with the STM tip after recording the ESCTM images is well identified in the EBSD inverse pole figure (IPF) and image quality (IQ) maps. Using the micro mark, the field of view of the ECSTM analysis can be easily located as indicated by the squares drawn in the EBSD maps.

The EBSD IPF map shows the crystallographic orientations of the grains in the analyzed local area (Figure 2(b)). The microstructure included in the ECSTM field of view consists of a single main grain color-coded in pink in the EBSD IPF map and labelled A in the ECSTM image. This main grain contains two types of sub-grains color-coded in green and blue in the EBSD IPF map and labelled B and C in the ECSTM image, respectively. The crystallographic orientation of main grain A is defined by the (-1 -2 4) plane parallel to the sheet plane, which is near the center of the IPF triangle. The orientation of the sub-grain B is (-26 -19 4), i.e. closer to the 101 pole, while that of the sub-grains C is (-11 21 -17), i.e. closer to the 111 pole. In the ECSTM image, the surface level for the sub-grains B and C is systematically measured to be lower than for grain A, suggesting a different surface reactivity of the sub-grains during surface preparation by electrochemical polishing (etching) with subsequent formation of the air-formed native oxide film followed by its *in situ* reduction.

The EBSD IQ map (Figure 2(c)) shows that the grain boundaries joining the main grains are of high angle random type (R GB) while nearly all of those joining the sub-grains with the main grains or the sub-grains themselves are of the CSL type. The length fraction of these CSL grain boundaries corresponds mostly to $\Sigma 3$ boundaries with a residual of $\Sigma 9$ boundaries. In the ECSTM field of view, only $\Sigma 3$ boundaries are included, joining the main grain A with either sub-grains B or sub-grains C.

In the ECSTM image of the topmost surface (Figure 2(a)), the $\Sigma 3$ grain boundaries can be located thanks to topography variations between the adjacent grains. Local sites at which the depth of the GB region and its variation after 2 and 4 CVs could be measured by line profile analysis are labelled from 1 to 9. The line profiles were obtained by averaging 30 adjacent line scans drawn across the grain boundary, covering a local distance of ~170 nm along the boundary. Typical profiles, measured at site 1 from the ECSTM images recorded after 0, 2 and 4 CVs, are superimposed in Figure 3(a). A first examination shows that the grain boundary region is well-marked by a depression at the surface between the two adjacent grains and that the depth of this gap varies after 2 and 4 CVs, suggesting reactivity. It should be emphasized that for several GBs observed in Figure 2(a), the line profile measured between the adjacent grains did not reveal any marked gap. This is related to the fact that Σ3 boundaries are joining the sub-grains B with main grain A and sub-grain $B_1$ with sub-grain $C_1$ as revealed in Figure 2(c) below and above the indented micro mark, respectively. This indicates that these Σ3 boundaries were inert, even during the electrochemical polishing pre-treatment where preferential pre-etching could be expected. This corrosion resistant behavior is as expected for Σ3 coherent twins (with a {111}-oriented GB plane) as previously observed by *in situ* ECSTM measurements of the initial stages of intergranular corrosion [37,38].

For the Σ3 GB line profiles along sites 1-9 in Figure 2(a), the depth of the intergranular region could be geometrically measured as the difference in height position between the bottom of the intergranular region and the surface level taken between the two adjacent grains. The results obtained after 0, 2 and 4 CVs are compiled in Figure 3(b). It is only after 4 CVs that a significant increase of the depth at the bottom of the GB region, indicative of irreversible dissolution and thus of the initiation of intergranular corrosion, could be observed at some of these GB sites. This is further discussed below. After 2 CVs, no net dissolution could be observed and the decrease of the depth of the GB region in some of the sites suggests that redeposition would dominate the local reactivity. Fluctuations between dissolution and redeposition have been shown to dominate the very early stages of the dissolution process, as directly observed by ECSTM at atomic step edges on single-crystal surfaces polarized at the onset of dissolution [30]. In the present case, a possible explanation is that the dissolution/redeposition process induced by the CV treatment might not be reversible in the GB region and that, in some of the measured sites, redeposition would dominate the local behavior at the bottom of the GB region where the GB plane terminates while dissolution would dominate in the immediate vicinity also part of the GB region.

The analyzed GB sites are classified as inert or reactive depending on their behavior after 4 CVs. For the GB sites 1-4 and 8, one observes no significant variation of the depth of the GB region, which indicates that no net dissolution occurred locally and these sites are therefore classified as inert. This corrosion resistant behavior is as expected for coherent twins [7,15,37,38]. In contrast, for the GB sites 5-7 and 9, the bottom of the GB region became deeper after 4 CVs, indicating that copper was irreversibly dissolved and that these sites are reactive. Being also of $\Sigma 3$ type, this reactive behavior suggests that these sites might include incoherent twins (i.e. $\Sigma 3$ boundaries with a GB plane markedly deviating from the exact {111} orientation) [7,15]. This $\Sigma 3$ twin-dependent corrosion resistant behavior is further discussed on the basis of complementary results presented hereafter.

At the reactive sites 5-7, the variation of the GB depth was measured to range between 1.2 to 2 nm. This is consistent with the equivalent thickness of copper (1.1 nm) that was estimated above to react at each CV assuming that 100% of the GB regions are reactive.

*3.3. Dissolution of random grain boundaries*

Figure 4 shows another local area of microcrystalline copper where ECSTM and EBSD analysis were combined. In this case, ECSTM images were recorded after 0, 2 and 6 dissolution/redeposition cycles. The EBSD maps indicate the micro marks made with the STM tip and that enable us to easily localize the STM field of view (marked by squares). The EBSD IPF map (Figure 4(b)) shows that the STM field of view includes six main grains of different orientations, labelled A to F, and three sub-grains, one in Grain D labelled G and two in Grain F labelled H and I. The ECSTM image (Figure 4(a)) is better resolved than the one in Figure 2(a) and reveals a terrace and step topography with different orientations of the step edges at the surface of Grains B, C and E. Such a topography is less marked at the surface of Grains A, D and F. The EBSD maps (Figure 4 (b) and (c)) enable the identification of the type of grain boundaries included in the STM field of view. The A/B, B/E, B/F, C/F, D/F and E/F grain boundaries are random boundaries (R GB) while all others are CSL type grain boundaries. Among the CSL boundaries, the A/E grain boundary is Σ9 and the A/D, D/E, D/G and F/H grain boundaries are Σ3 boundaries.

Similarly to Figure 2 (a)), the local sites between adjacent grains, where the depth of the GB region and its variation after 2 and 6 CVs, were measured by line profile analysis, are labelled from 1 to 12. The results are compiled in Figure 5 and Table 1. First, the behavior of the grain boundaries of random type measured at Sites 4, 5 and 7 corresponding to the B/F, C/F and A/B interfaces, respectively, is examined. In all three cases, the depth of the GB region was determined to significantly increase after 6 CVs with depth increase values ranging from 0.7 to 0.9 nm depending on site, indicating irreversible dissolution. For two sites (Site 7 could not be measured conclusively), irreversible dissolution was even observed after 2 CVs indicating their reactivity in the very first stages of dissolution induced by the CV treatments. Clearly, these random type grain boundaries are reactive and can be classified as susceptible to the initiation of intergranular corrosion, in agreement with other studies on the propagation of intergranular corrosion [6,8,10,13-15,17].

*3.4. Effect of deviation from CSL orientation*

Similarly to the data in Figure 2, the analysis of the behavior of the Σ3 grain boundaries in Figure 4 shows that they can be potentially classified as either susceptible or resistant to the initiation of intergranular corrosion. At Sites 1-3 and 11, along the A/D grain boundary, a significant increase of the depth of the GB region was measured after 6 CVs (Figure 5, Table 1), indicating the susceptibility of this Σ3 boundary to the initiation of intergranular corrosion. In contrast, at Sites 6 and 10 along the D/E grain boundary, also of the Σ3 type, no significant increase of the depth of the GB region could be measured, indicating a corrosion resistant behavior. This is also the case at Site 12 along the D/G grain boundary of Σ3 type and at Site 9 where a Σ3 type grain boundary was visualized by ECSTM.

At Site 8 in Figure 4, corresponding to a Σ9 grain boundary at the A/E interface as revealed by the local EBSD analysis (Table 1), no significant variation in depth of the GB region was measured after 2 CVs but a significant depth increase was observed (1.2 nm) after 6 CVs (Figure 5). This shows that this type of CSL boundary may reversibly react in the very first stages of dissolution but still can be classified as susceptible to the initiation of intergranular corrosion. To our knowledge, this type of behavior is observed for the first time at the nanoscale.

Owing to their geometry, coherent Σ3 grain boundaries are assigned the following characteristics: no intrinsic defect, no potential segregation and no intergranular corrosion [51]. So all Σ3 grain boundaries showing reactivity to dissolution should be incoherent due to their deviation with respect to the ideal GB plane or no longer considered Σ3 grain boundaries due to their misorientation angle.

In order to better rationalize the data, the EBSD analysis was refined by extracting the GB misorientation angle $\theta$ and deviation angle $\varphi$ with respect to the ideal GB plane at the local Sites 1-12 in Figure 4 where the ECSTM analysis was performed. The results are compiled in Table 1. The Brandon criterion [52] ($\Delta\theta\ max\ =\ 15°\ \Sigma^{-1/2}$) gives a tolerance of 8.67° for the Σ3 misorientation, which means that grain boundaries with a misorientation angle outside the range (60±8.67)° can no longer be considered Σ3. The Palumbo-Aust criterion [53] ($\Delta\theta\ max\ =\ 15°\ \Sigma^{-5/6}$) gives a tolerance of 6°. In our case, all the Σ3 grain boundaries listed in Table 1 respect both criteria criterion and can be considered as Σ3 according to their misorientation angle. Their different behavior is then discussed based on the deviation angle of the GB plane (i.e. incoherent vs. coherent twins).

At Sites 1-3, which were measured reactive by ECSTM, the Σ3 boundaries have their planes slightly deviating from the ideal {111} orientation. Indeed, the GB deviation angle is 1.7° at best instead of 0°. At Site 11 also measured reactive by ECSTM, the local EBSD analysis did not allow to retrieve reliable data. However, it is likely that the local misorientation is similar to that measured at Sites 1-3 since Site 11 is located along the same boundary. This clearly demonstrates that the grain boundary at the A/D interface can be considered as an incoherent twin containing a step structure accommodating the deviation of the boundary plane from the exact CSL plane and that this step structure makes it susceptible to the initiation of intergranular corrosion at the nanoscale.

At Sites 6 and 10 measured inert by ECSTM, the Σ3 boundaries have deviation angles of 1 and 0.8°, respectively, closer to those of the exact Σ3 CSL. This is confirmed at Sites 9 and 12 where the deviation angle is 0.9 and 0.6°, respectively. Hence, it can be concluded that the grain boundaries at these sites resist nanoscale intergranular corrosion because they are close to perfect Σ3 coherent twins. Furthermore, these data show that a transition from an inert to reactive behavior in the initiation of intergranular corrosion can be expected with increasing deviation from the exact orientation of the CSL plane. For a Σ3 twin, this transition would occur for a deviation angle between 1° and 1.7° owing to the increased density of step regions introduced in the boundary plane with increasing deviation.

Models of coherent and incoherent Σ3 twins are presented in Figure 6. With increasing angle $\varphi$ of the GB plane with respect to the twinning plane, the linear density of steps in the GB plane $\rho_{step}$ increases according to:

$$\rho_{step} = \frac{1}{L} = \frac{d}{\tan \varphi} \qquad \text{Eq. (2)}$$

where d is the reticular distance of the (111) planes. For deviation angles of 1° and 1.7°, the linear density of steps is 11.9 and 7 nm$^{-1}$, respectively, showing the precision needed for the twins to remain immune to the initiation of intergranular corrosion.

For the Σ9 boundary measured at Site 8, the GB misorientation angle (38.7°) also satisfies the Brandon (38.94 ± 5°) and Palumbo-Aust (38.94 ± 1.4°) criteria. The deviation angle (0.6°) is close to that expected for an ideal Σ9 CSL (0°). Still, this boundary was measured reactive by ECSTM. This shows that the IG initiation corrosion behavior of these higher CSL boundaries does not tolerate as much deviation from the exact CSL plane as for Σ3 twins.

# 4. Conclusions

Combined ECSTM and EBSD analysis of the same local area was applied for the first time to study at the nanoscale the initiation of intergranular corrosion for various types of grain boundaries. The adopted methodology included *in situ* ECSTM local surface analysis before and after electrochemically-induced dissolution/redeposition in quantities not exceeding a few equivalent monolayers. This was followed by micro marking of the surface by indenting the surface with the STM tip and subsequent repositioning of the EBSD analysis in the micro marked local area. Data analysis included the determination of the depth of the intergranular region from line profiles of the topmost surface and correlation with the grain boundary character data obtained by EBSD analysis at the same sites.

The results obtained on microcrystalline copper in the active state of dissolution in 1 mM HCl(aq) aqueous solution show that the initiation of intergranular corrosion measured at the nanoscale is dependent on the grain boundary character. Among the boundaries that could be locally analyzed with the combined ECSTM/EBSD approach, random high angle boundaries were found susceptible to the nanoscale initiation of localized corrosion as well as $\Sigma 9$ coincidence boundaries with a deviation of 0.6° from the exact orientation of the GB plane. For $\Sigma 3$ coincidence boundaries, the behavior was found dependent on the deviation of the GB plane from the exact CSL orientation. Significant initiation of dissolution was observed for a deviation angle of 1.7° and above while corrosion resistance was observed for a deviation angle of 1° or less, indicating the existence of a transition from immunity to sensitivity for a deviation angle between 1° and 1.7° and a corresponding linear density of steps (i.e. misorientation dislocations) in the GB plane between 11.9 and 7 $nm^{-1}$, respectively.

**Acknowledgments**

This project has received funding from the European Research Council (ERC) under the European Union's Horizon 2020 research and innovation programme (Advanced Grant agreement No 741123).

# Tables

Table 1 Grain boundary crystallographic characteristics as measured by EBSD at the sites marked in Figure 4 and variation of the depth of the GB region and IGC behavior as measured by ECSTM after 6 CVs

| Site | GB type | GB misorientation angle $\theta$ (°) | Deviation angle $\varphi$ (°) | Variation of GB depth (nm) | IGC behavior |
|---|---|---|---|---|---|
| 4 | R | 55.3 | - | 0.8 | Reactive |
| 5 | R | 50 | - | 0.9 | Reactive |
| 7 | R | 54.6 | - | 0.7 | Reactive |
| 8 | Σ9 | 38.7 | 0.6 | 1.2 | Reactive |
| 1 | Σ3 | 58.7 | 1.7 | 0.8 | Reactive |
| 2 | Σ3 | 59 | 2.1 | 0.7 | Reactive |
| 3 | Σ3 | 58 | 2.3 | 0.7 | Reactive |
| 11 | - | - | - | 0.8 | Reactive |
| 6 | Σ3 | 59.5 | 1.0 | ~0 | Inert |
| 10 | Σ3 | 59.6 | 0.8 | ~0 | Inert |
| 9 | Σ3 | 59.7 | 0.9 | ~0 | Inert |
| 12 | Σ3 | 59.8 | 0.6 | ~0 | Inert |

**Figure captions**

Figure 1 Cyclic voltammograms in the ECSTM cell for microcrystalline copper in 1 mM HCl(aq), scan rate = 20 mV/s: (a) CV recorded between anodic dissolution and hydrogen evolution. (b) Zoom in the anodic dissolution region.

Figure 2 Microcrystalline copper as first imaged *in situ* by ECSTM and then mapped by EBSD after repositioning in the same local area: (a) Topographic ECSTM image of the initial metallic state at E = -0.55 V/SHE in 1 mM HCl(aq) (Z range $\Delta Z = 24$ nm, tip potential $E_{tip}$ = -0.4 V/SHE, tunneling current $I_t$ = 2 nA); (b) EBSD inverse pole figure (IPF) map; (c) EBSD image quality (IQ) map. In (a), the grain A and the sub-grains B and C are labelled as well as the local GB sites selected for depth profile analysis. In (b) and (c), the field of view analyzed by STM is marked by a square. In (c), black lines denote the random grains boundaries (R GB), red lines the $\Sigma 3$ twin boundaries and green lines the $\Sigma 9$ boundaries.

Figure 3 Assessment of the GB reactivity from the ECSTM data after applying 0, 2 and 4 CVs: Average topographic line profiles measured across the GB at site # 1 in Figure 2(a) (a); Bar graph of the depth measured across the grain boundaries at the sites # 1 to 9 in Figure 2(a).

Figure 4 Local area of microcrystalline copper first imaged *in situ* by ECSTM and then mapped by EBSD after repositioning: (a) Topographic ECSTM image of the initial metallic state at E = -0.55 V/SHE in 1 mM HCl(aq) ($\Delta Z$ = 12 nm, $E_{tip}$ = -0.4 V/SHE, $I_t$ = 1.8 nA); (b) EBSD IPF map; and (c) EBSD IQ map. In (a), the grains are labelled from A to H and local GB sites selected for depth profile analysis are labelled from 1 to 12. In (b) and (c), the square marks the STM field of view. In (c), black lines denote the random grains boundaries (R GB), red lines the $\Sigma 3$ boundaries and green lines the $\Sigma 9$ boundaries.

Figure 5 Bar graph of the depth measured across the random and CSL grain boundaries at the sites # 1 to 12 in Figure 4(a) after applying 0, 2 and 6 dissolution cycles.

Figure 6 Models of (a) coherent and (b) incoherent $\Sigma 3$ twin boundaries. <111> is the rotation axis common to both crystals and $\theta$ the rotation angle. $\varphi$ is the deviation angle of the GB plane from the twinning plane. Steps are introduced on the GB plane for increasing values of $\varphi$.

**Figure 1**

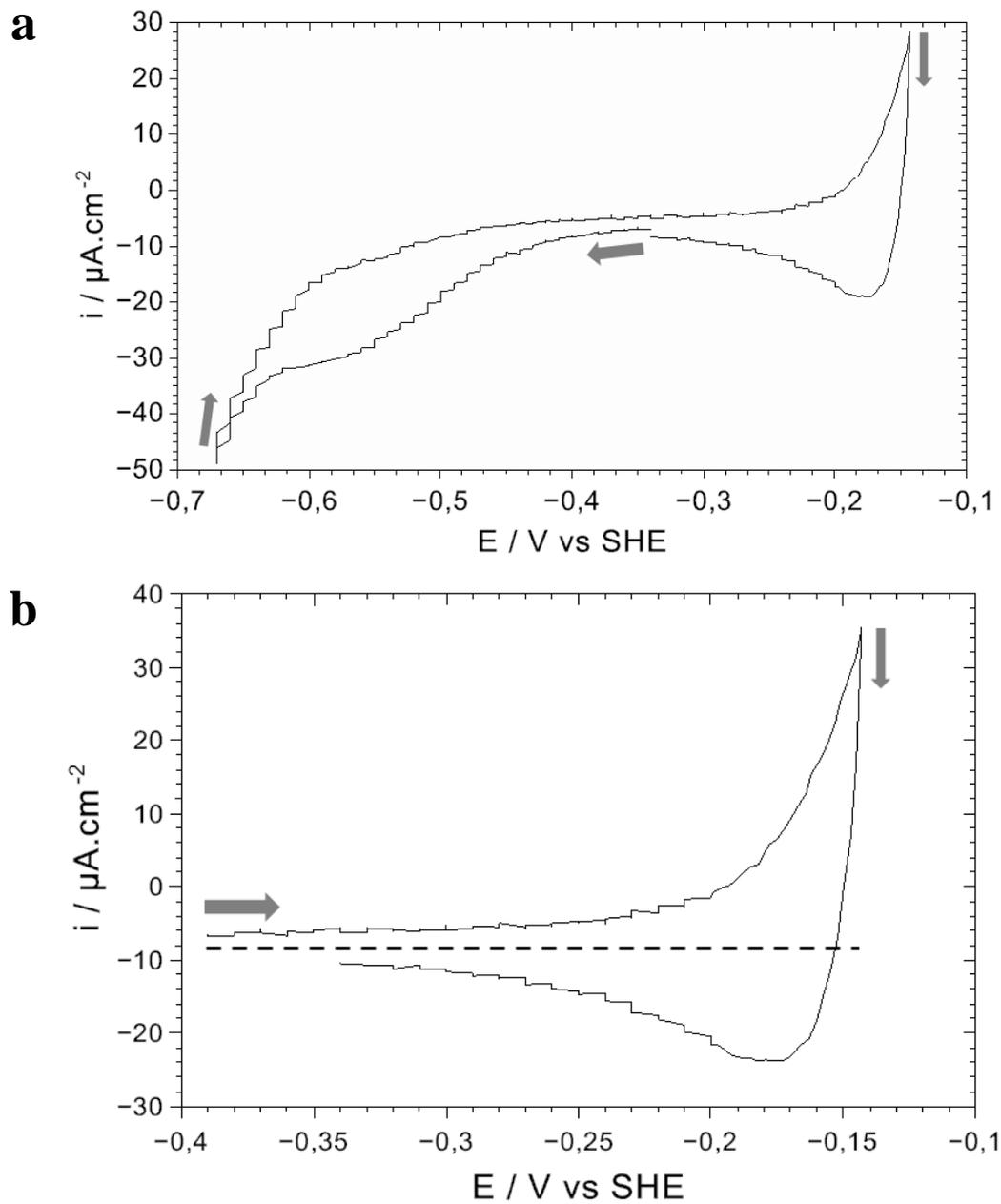

**Figure 2**

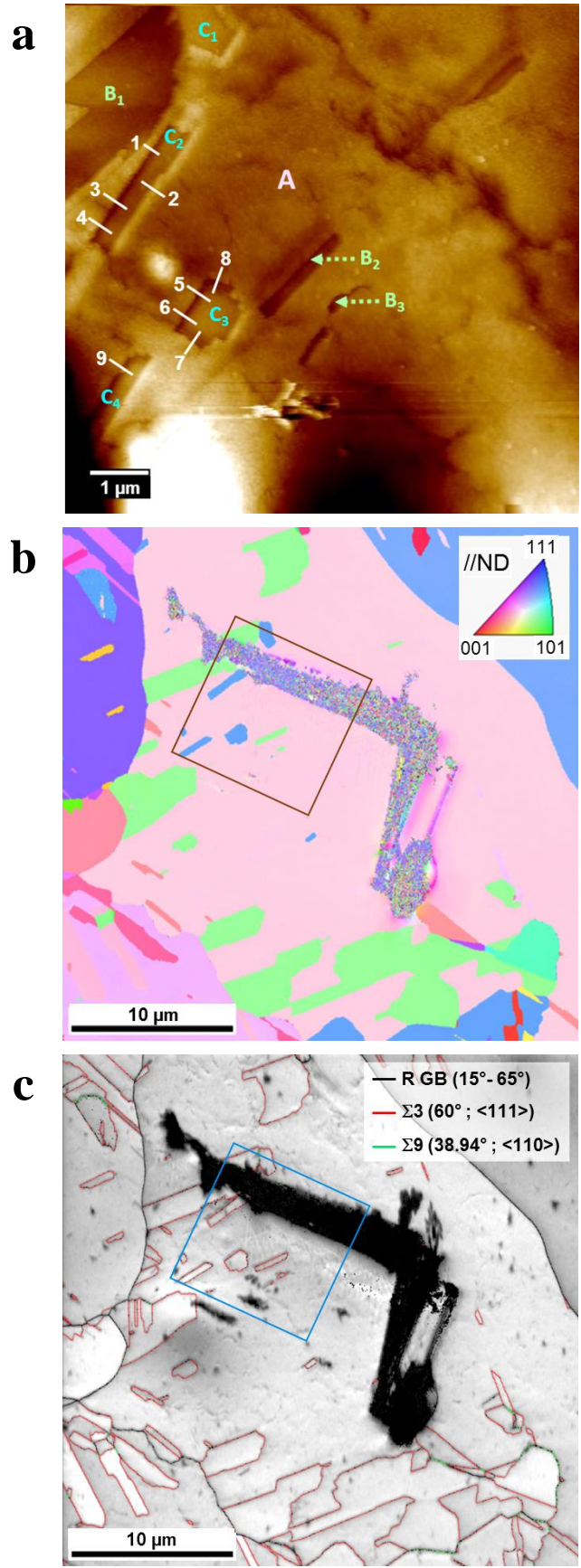

**Figure 3**

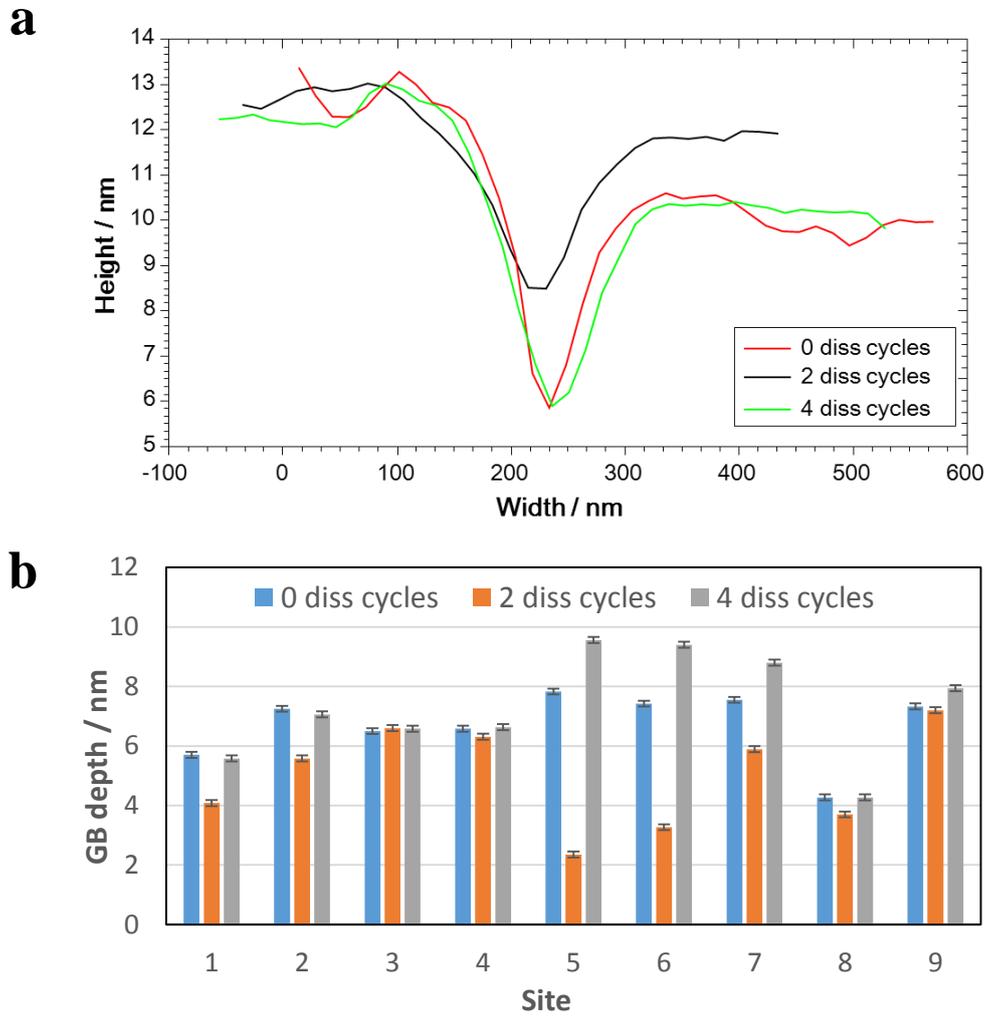

**Figure 4**

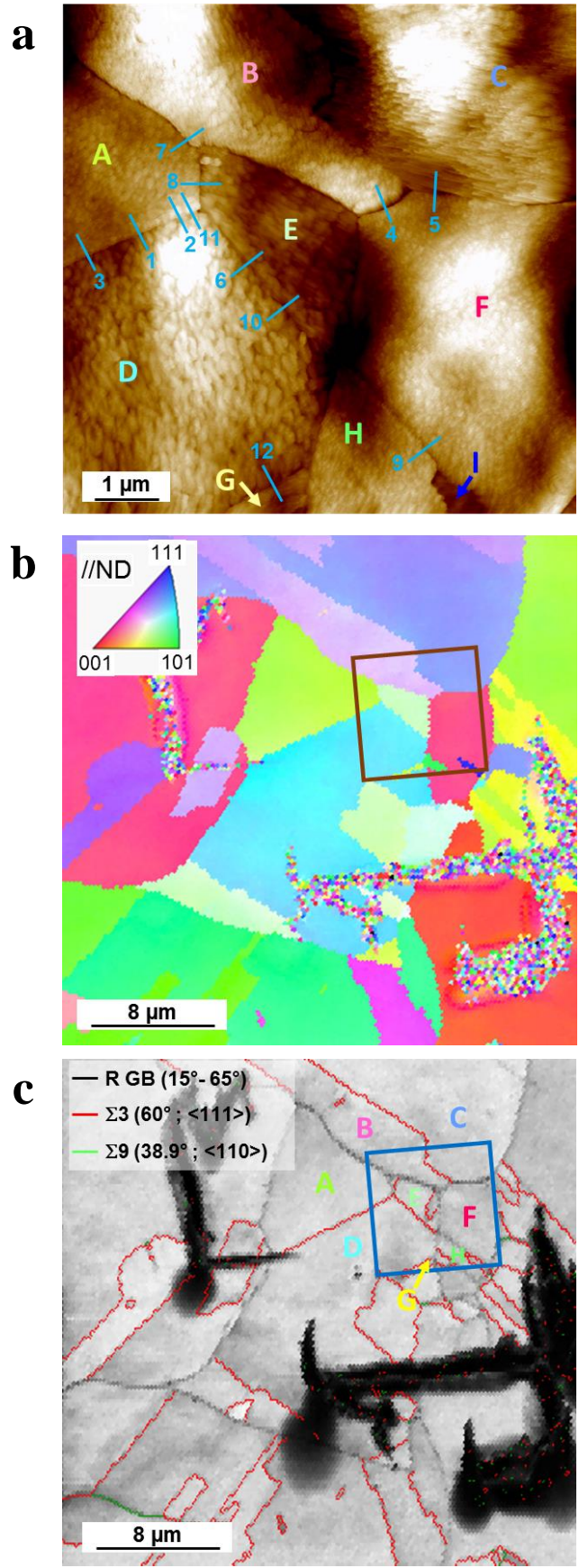

**Figure 5**

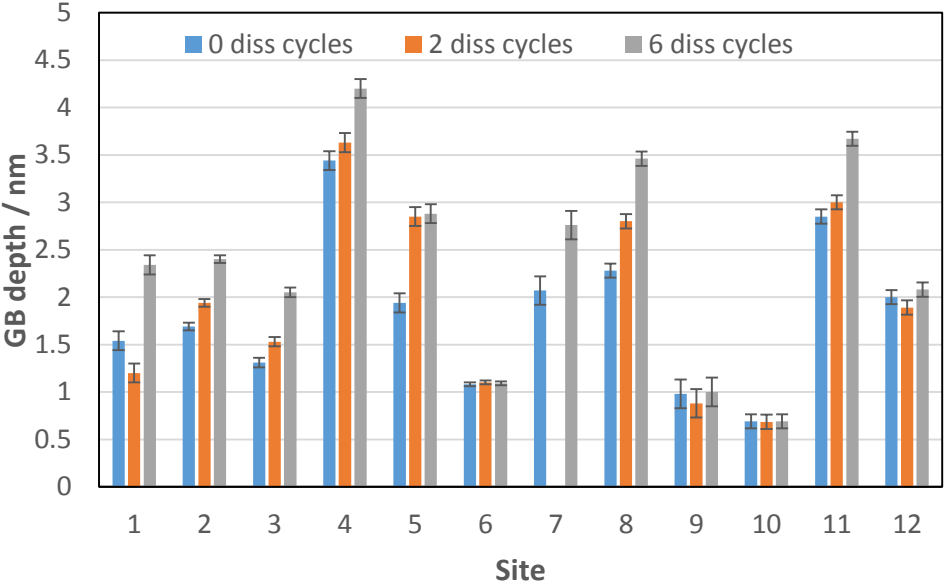

**Figure 6**

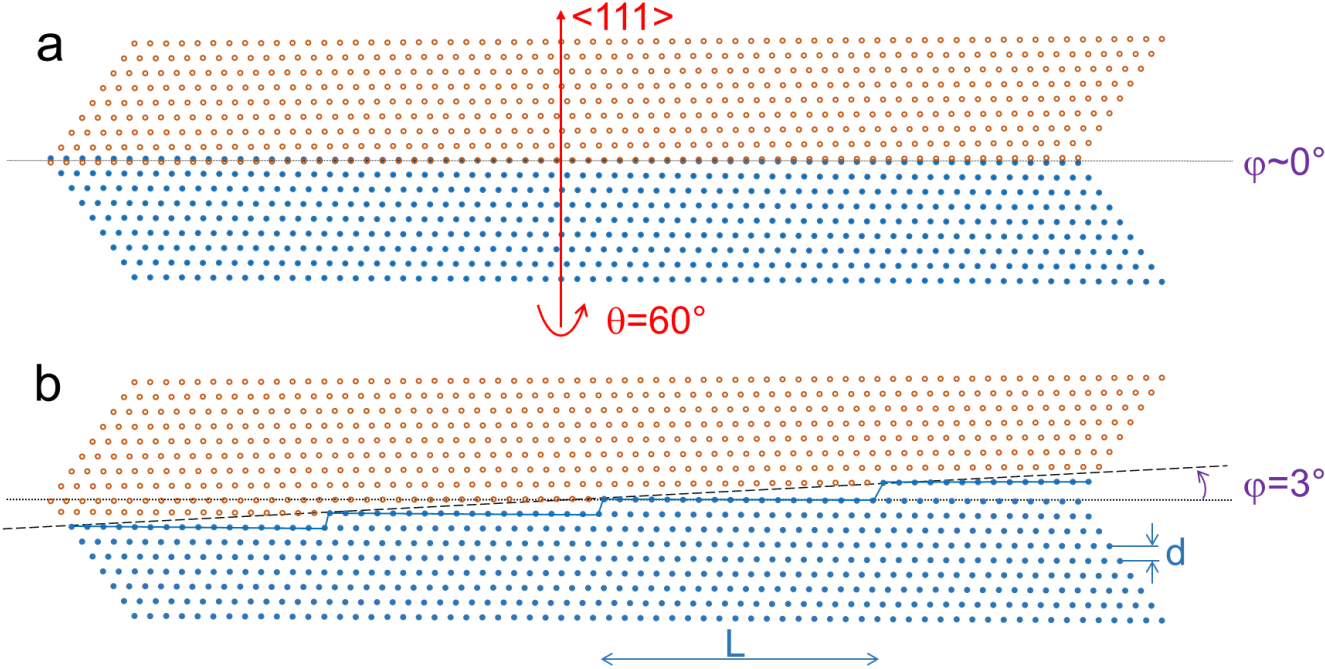